\documentclass{article}
\def\beq{\begin{equation}} 
\def\eeq{\end{equation}} 
\def\bea{\begin{eqnarray}}  
\def\eea{\end{eqnarray}}
\def\bi{\begin{itemize}}  
\def\ei{\end{itemize}}  
\def\beqa{\begin{eqnarray}}  
\def\eeqa{\end{eqnarray}}

\def\pa{\partial}

\def\r2{\sqrt{2}}

\def\nn{\nonumber \\}

\def\ca{{\cal A}}  

\def\mg{m_{3/2}}

\begin{document}
\thispagestyle{empty}

\begin{flushright}   CERN-TH/2001-244
\end{flushright}
\vskip 2cm
\begin{center}
{\huge Low Energy  Supersymmetry in Warped Brane Worlds}
\vspace*{5mm} \vspace*{1cm} 
\end{center}
\vspace*{5mm} \noindent
\vskip 0.5cm
\centerline{\bf Zygmunt Lalak}
\vskip 1cm
\centerline{\em Theory Division, CERN}
\centerline{\em CH-1211 Geneva 23}
\vskip 0.3cm
\centerline{\em Institute of Theoretical Physics}
\centerline{\em University of Warsaw, Poland}
\vskip2cm

\centerline{\bf Abstract}
\vskip .3cm
We discuss physical implications of the  four-dimensional 
effective supergravity, that describes low-energy physics 
of the Randall--Sundrum model with moduli fields 
in the bulk and charged chiral matter living on the branes.
Cosmological constant can be cancelled through the introduction of a 
brane Polonyi field and a brane superpotential for the 4d dilaton.
We deduce a generalization of the effective 4d action to 
the case of a general, not necessarily exponential, warp factor.
We note, that breakdown of supersymmetry in generic warped models 
may lead to the stabilization of the interbrane distance.
\vskip1cm
\begin{flushleft}   
Talk given at SUSY'01, June 11-17 2001, Dubna, Russia.\\  
\vskip0.5cm
September 2001
\end{flushleft}
\newpage


Brane worlds with warped geometries offer new perspectives in 
understanding the hierarchy of mass scales in field theory models 
\cite{lala-rs1,lala-rs2,lala-rs3}. The initial hope was that the mere presence of extra 
dimensions 
would be a natural tool to control mass scales in gauge theories 
coupled to gravity. However, the realization of these simple ideas 
in terms of consistent models eventually called for quite 
sophisticated constructions, such as the brane--bulk supersymmetry that 
has been constructed in \cite{lala-bagger,lala-gp,lala-hiszp,lala-flp,lala-flp2,lala-flp3,lala-kallosh,lala-zucker}. \\
In these models 
four-dimensional hypersurfaces (branes) hosting familiar gauge and charged matter fields, are embedded in a five-dimensional ambient space, the bulk, populated 
by gravitational and gauge-neutral fields. The bulk degrees of freedom 
couple to the fields living on branes through various types of interactions. 
Some of these interactions are analogs of an interaction between 
electromagnetic potential and charge density located on branes -- this is the case of the fields that are $Z_2$-even on an $S^1/Z_2$ orbifold forming the 
fifth dimension; some of them are rather analogs of the derivative coupling of the potential to the electric dipole moment density located on branes -- 
that is the case for the interactions of the $Z_2$-odd fields on $S^1/Z_2$. 
These interactions lead to the formation of nontrivial vacuum 
configurations in the brane system. In particular, the solutions 
of Einstein equations of the form $ds^2 = a^{2}(x^5) ds^{2}_{4} + b^2(x^5) (dx^5)^2$ are usually allowed, as in the original Randall--Sundrum (RS)
models, where $ds^{2}_{4}$ is the Minkowski, anti-deSitter or deSitter 
metric in 4d, and $a(x^5)$ is a warp factor. 

We find it interesting to study  
the consequences of nontrivial warp factors for the low energy 
physics. To this end one needs a class of realistic brane models with chiral matter, gauge fields and gravity and other bulk fields consistently coupled 
to each other. We shall use the locally supersymmetric brane
models in 5d which have been constructed in \cite{lala-flp,lala-flp2,lala-flp3}. These 
are the only existing examples of consistent supergravity--brane matter models,
which allow to watch a delicate interplay of local supersymmetry and warped geometry. In the framework of this theory 
it is possible to make precise correspondence between the five-dimensional brane world picture and various aspects of the four-diemnsional effective theory in the infrared regime. 
To study the phenomenology of the models in detail, and in particular to demonstrate how cosmological constant can be cancelled, we turn again to the Randall--Sundrum case where the calculations can be done expliciltly. 
Later on we argue that one can 
generalize results summarized here    
 for the Randall-Sundrum background to the case of an arbitrary 
warp factor which arises as a BPS solution of 5d supergravity with branes. 
We discuss the possible form of low-energy supergravity in the general case. 
In particular, we argue that the appearance 
of a $T$--dependent superpotential is a general phenomenon, which in many 
cases leads to the stabilization of the radion field 
(hence to stabilization of the extra dimension through the supersymmetric version of the Goldberger--Wise mechanism). 
We shall try to establish relations between supersymmetry (and its breakdown),
stabilization of the radion, vanishing cosmological constant and the form 
of low energy theory in the infrared. 

In \cite{lala-flp3} we found that the low energy effective theory of the zero mode fluctuations in the supersymmetric Randall-Sundrum scenario can be described by N=1 supergravity with the Kahler potential and the superpotential given by:
 \bea &
K=-M^{2}_P \log(S + \bar{S}) - M^{2}_P \log \left ( 1-e^{-(T+\bar{T}-\frac{k}{3 M^3} |\Phi_2|^2)} -\frac{k}{3 M^3} |\Phi_1|^2   \right )
&\nn&
W= 2 \sqrt{2} (W_1 + e^{-3 T} W_2). 
\eea
In  the above, $T$ is the radion multiplet: $T= (R_0 + i \r2 \ca_5) k\pi\rho$ ; its real part describes the fluctuations of the distance between the two branes. The dilaton multiplet $S=V_0 + i \sigma$ originates from the universal hypermultiplet present in the 5d set-up. The multiplets $\Phi_1$ and $\Phi_2$ originate from the chiral matter multiplets confined to the hidden and the visible branes, respectively. The superpotential $W$ is constructed out of the brane superpotentials; the contribution from the warped (visible) brane is suppressed,
at the level of the Lagrangian, by an exponent of the radion field.   

The basic features of the 4d effective supergravity are as follows. If the superpotentials $W_i$ have vanishing expectation values, the vacuum solution is just 4d Minkowski space, supersymmetry remains unbroken, and all moduli are massless. 
 As soon as the superpotentials get non-trivial vacuum expectation values, supersymmetry is broken and a potential for the T modulus is generated. The radion gets stabilized at the value: $e^{k\pi\rho (R_0 + i3\r2\ca_5)}=-\frac{W_2}{W_1}$ and gets a mass of the order of the supersymmetry breaking scale. The gravitino mass is $m_{3/2} \approx \sqrt{2/V_0}\frac{|W_1|}{M_P^2}$ and supersymmetry is broken by $F_S$ ( $F_T$ remains zero). 

Fixing $R_0$ at a value corresponding to the scale of $1\;TeV$ at the warped brane demands a fine tuning between the values of $W_1$ and $W_2$. 
In the proposal \cite{lala-gw} by Goldberger and Wise it was noticed, that such a tuning may be significantly softened by the ratio of the $k^2$ and the mass $M^{2}_{\phi}$
of the GW bulk field, present in their formula for $\langle R_0 \rangle$. 
We note, that in our model, where we require a supersymmetric bulk Lagrangian,
the mass of the field playing the role of the GW scalar is fixed by supersymmetry to be $M^{2}_{\phi}= - 3 k^2$. Hence, there is no free modulation factor, 
and we need a tuning between the boundary values of the relevant scalar field
(which are $W_i$) to stabilize the radion at the desired scale.  

To shed some light on the relation between stabilization of $T$ and 
vanishing of $F^T$, let us decouple the universal hypermultiplet from 
the model. Then the K\"ahler potential is $K=-3 \log (1 - e^{-(T + \bar{T})} )$
and the scalar potential takes the form 
\beq
V= \frac{3}{(1 - e^{-(T + \bar{T})} )^3} \left ( |e^{-2 T} W_2 + e^{-\bar{T}}W_1|^2 - | W_1 + e^{-3 T} W_2|^2 \right ).
\eeq
The potential does not depend on the  imaginary part of $T$, but the real 
part of $T$ is fixed: there is a minimum of the potential at $e^{- 2 Re(T)} =  
\frac{|W_1|^2}{|W_2|^2}$. However the expectation value of $F^T$ doesn't need to be zero
\beq
|F^T|^2 \approx \frac{2 |W_1|^4}{|W_2|^2} (1 + \cos ( \phi_2 -\phi_1 - 3 Im (T))),
\eeq
where $\phi_i$ are the phases of $W_i$. The above expression vanishes for 
discrete  values of the imaginary part of $T$, but there is a continuum of supersymmetry breaking vacua. This should be compared to the case with the 
dilaton $S$ present. As long as the superpotential is independent of $S$, the 
effect of the dilaton dependence of the K\"ahler potential consists in a
partial cancellation of the term $-3 e^K |W|^2$ in the scalar potential: effectively the coefficient of this term becomes $-2$. The result of that change is that $Im(T)$ is 
no longer a flat direction of the potential, and the minimum corresponds to 
a value of $Im(T)$ that gives $|F^T|^2=0$.  

The main drawback of this scenario  is that  a large, negative cosmological constant is generated $V_{vac} \approx -m_{3/2}^2 M_p^2$ and thus the model as it stands has no direct applications for phenomenology. 

The first guess would be that the cosmological constant can be cancelled (possibly with the need of a fine-tuning) by an addition of a contribution from the boundary matter sector. Indeed, if allow the superpotentials $W_i$ to depend on the scalar $\Phi_i$, the brane potentials in the 5d set-up obtain the 
contribution: $V_b=\frac{1}{2 V}\delta(x^5) 
 |\frac{\pa W_1}{\pa \Phi_1}|^2 +   
\frac{1}{2V}\delta(x^5-\pi\rho)  |\frac{\pa W_2}{\pa \Phi_2}|^2$  which is always positive, and in principle might cancel the negative vacuum energy. However, this expectation is not fullfilled.
It turns out, that to cancel the cosmological constant one needs a combination
of a brane potential (we shall simply take a Polonyi field on the Planck brane), and a potential for the second bulk scalar, the 4d dilaton $S$. As the concrete
realization of the second piece we simply take two gaugino condensates on the 
visible brane. An example of the superpotential that works this way is 
\beq
W= W_1(\Phi_1) + W_2 (S) e^{-3T}.
\eeq  
The resulting model works similarly to the 4d racetrack models - stabilizes
the dilaton and the radion. In addition, 
it is possible now to tune the cosmological constant to an arbitrarily small 
positive or negative value. 
The radion $R_0$ is  stabilized around the value $e^{k R_0 \pi \rho} \sim |\frac{W_2}{W_1}|$ and its physical mass is of the order of the supersymmetry breaking scale: $m_R \sim m_{3/2}$. The vacuum value of the T multiplet pseudoscalar superpartner $\ca_5$ is set by a phase difference  of the superpotentials: $W_1 \bar{W_2}e^{i 3 \r2 \ca_5 \pi \rho}= -|W_1||W_2|$ and its mass is of the same order as that of the radion. The dilaton $V_0$ is also 
stabilized and
the vacuum value of the axion $\sigma$ is zero. 
\vskip0.2cm
The RS2 model with two branes is often said to offer a solution 
to the perturbative hierarchy problem. The basic observation is that 
on the warped brane all the mass scales initially present in the Lagrangian
become universally scaled down by the factor $a(\pi \rho)$: $m \rightarrow 
a( \pi \rho) m$, so that on that brane the local reference scale becomes 
$\mu = M a(\pi \rho)$ rather than the 5d gravitational scale $M$. 
Moreover, the masses of the Kaluza--Klein towers of the bulk fields start with 
$m_{KK}=\mu$ and the coupling of heavy KK modes to matter localized on the 
warped brane is set by $\mu^{-1}$. This means, that at energies $E$ 
somewhat larger than $\mu$ the KK modes of gravitons and bulk moduli 
couple strongly to the brane fields, and the whole mixture needs to be 
treated as a single model, unfortunately -- strongly coupled in the 
(super)gravity description. One can argue on the basis of the conjectured 
gravity--CFT theory that the upper scale of perturbative physics is slightly 
higher than $\mu$, $\Lambda_{cut-off} = \mu (M/k)$, where $c=(M/k)^3$ has the interpretation of the central charge of the conjectured dual CFT, so that 
there is a region where the 5d weakly coupled bulk-brane supergravity is 
a good description of physics. As to the question, whether in the presence of the cut-off one needs supersymmetry, one should notice that one still needs to 
control quantum corrections in the vicinity of the unwarped brane, and in 
particular one needs to protect the results of the stabilization of the 
radion, the mechanism of which {\em is not} confined to the warped brane. 
As we saw, once supersymmetry breakdown is tiggered on a brane or in the bulk, the vacuum configuration of scalar fields breaks supersymmetry everywhere
in the system. Thus the effects of the breakdown are not confined to any 
particular region in space, but just to the contrary, they get spread over all 
sectors of the model by the mechanism which is from the 4d point of view a mixture of gravity and anomaly mediation. 
Beyond this, if there is the underlying string theory behind the warped 
gravity, then supersymmetry inherited from the string should be visible in 
the gravity--picture description of the system. 

Let us review then the mass scale patterns  that are possible 
 in the weakly couplead brane-bulk supergravity regime. 
First of all, the gravitational mass 
splittings in the multiplets living on the branes are set by $\mg 
\sim \frac{|W_1|}{M^{2}_P} = \frac{a (\pi \rho) |W_2|}{M^{2}_P}$. 
Now, we would call $|W_2|$ perturbative, if 
$|W_2| \leq \mu^3$ ($\mu$ is the upper scale of perturbativity)
If we insist that $\mu = 1 \; TeV$, then we obtain 
$\mg \leq 10^{-45} \; TeV$. This cannot be completely excluded, but to 
obtain a realistic mass split in the multiplets one would need to invoke some 
additional mechanism of supersymmetry breakdown mediation, like gauge 
mediation, which makes the whole scheme complicated. 
The other possibility is to say that $\mg \approx \mu = 1 \; TeV$,
and to find out that $|W_2| = M^{3}_P \gg \mu^3$. It is easy to see that 
in such a case $|W_1| = (10^{13} \; GeV)^3$. Hence, on this case one has 
a set of three scales in the perturbative sector, $M_2 = 1 \;TeV$, $M_1 = 
10^{13} \; TeV$ and $M_P$, two of which are beyond the naive range of perturbativity. 
A different option is possible. Let us request that $\mg = 1 \; TeV$ and 
$W_2$ be perturbative. Then we obtain the needed value of the 
warp factor $a(\pi \rho) \approx 10^{-4}$. One finds that $M_1 \approx 10^{14} \; GeV$ and $M_2 \approx 10^{-1} M_1$. The scales of the brane physics 
are within the perturbative regime, required tuning between branes rather 
mild, and the scale $\mu$ is pretty close to 
the usual GUT scale. This might suggest, that actualy it is more 
natural in the supersymmetric Randall--Sundrum to think of the warped 
brane as of a $M_{GUT}$-brane, rather than a $TeV$ brane; with supersymmetry 
required anyway, there is simply no obvious preference for the $TeV$ version. 
%
\vskip0.3cm
Having established the form of the effective supergravity below 
$\Lambda = c^{1/3} \mu$, it is easy to compute the one-loop contribution 
to the effective Lagrangian, hence also the one-loop contribution to the 
vacuum energy coming from this momentum range. The general result is that 
of a cut-off 4d supergravity
\beqa
&L^{(1-loop)} = e \frac{1}{32 \pi^2} \frac{\Lambda^2}{M^{2}_P} (2 (N-1) (\mg^2 M^{2}_P + V^{(0)} ) - 4 \mg^2 M^{2}_P & \nonumber \\
&- \frac{N}{2} R - 2 mg^2 \langle R_{i \bar{j}}
G^i G^{\bar{j}} \rangle )
+ O(\log(\Lambda^2 / M^{2}_P) \mg^4) + {\rm derivative\;\;terms},&
\eeqa
where a maximally symmetric 4d gravitational background is assumed, and 
$R_{i \bar{j}}$ is the Ricci tensor of the K\"ahler manifold spanned by chiral multiplets. The leading contribution to the cosmological constant is 
\beqa 
&{\cal V}_{vac} = V_0 + \frac{1}{32 \pi^2}\frac{\Lambda^2}{M^{2}_P}(-
2 (N-1) \mg^2 M^{2}_P - 22 V_0  + 2 mg^2 \langle R_{i \bar{j}}
G^i G^{\bar{j}} \rangle)& \nn \\
& + O(\log(\Lambda^2 / M^{2}_P) \mg^4).&
\eeqa
In the model without the superpotential for the 4d dilaton we have arrived 
at the $AdS_4$ solution with $V_0 = - 2 \mg^2 M^{2}_P$. 
This obviously cannot be compensated by the one--loop correction, since 
the leading positive contribution of the latter is of the order $\mg^2 \Lambda^2$. Nevetheless, if one tries the first scenario, with nongravitational mediation, then one finds $V_0 \sim (-1) o(10^{-120}) M^4_P$, and the loop correction
shall be $a^2(\pi \rho)$ ($10^{-30}$) times smaller in magnitude. 
These orders of magnitude of the vacuum energy fall into the domain of 
the quintessence physics (except that the sign we originally have 
is unsuitable), and we shall not comment on this issue any further.  
In the cases with a $TeV$ gravitino mass the magnitude of the vacuum energy 
becomes disastrously large. W have shown, however, that once we switch on an
effective superpotential for $S$, the situation changes, and it becomes possible to tune the tree-level vacuum energy to zero (we shall discuss details of 
how does one achieve the cancellation later on); in this case the 
leading contribution to vacuum energy is the one arising at one-loop order. 
The order of magnitude of this contribution is 
\beq |{\cal V}|_{vac} \sim \frac{1}{16 \pi^2} a^2 (\pi \rho) \mg^2 M^{2}_P,
\eeq
which for the $TeV$ gravitino is $\sim a^2(\pi \rho)\; 10^{-30}\; M^{4}_P$ 
(times the loop factor). For both cases, the one of the $TeV$ warped brane and the one more natural one of the $M_{GUT}$ scale warped brane, the resulting numbers are large when compared to existing bounds. The message from this 
discussion is such, that indeed, in the case of warped brane worlds 
supersymmetry protects the scale of vacuum energy, but only if one makes 
the gravitino mass sufficiently small. Firstly, this is a very fine tuning, 
secondly -- on the basis of numbers we have put into the formulae we 
conclude, that this requires a nongravitational mechanism of supersymmetry 
breaking transmission to the observable supermultiplets. Thus curing one problem, one generates another one, which is as much stumbling. 
%
  
We have just argued that it is possible, by creating a brane 
superpotential for the superfield $S$, to achieve the stabilization of 
the moduli space while keeping the tree-level cosmological constant zero. 
To summarize, to achieve such a situation we had to introduce yet another,
beyond moduli $S$ and $T$, field participating actively in supersymmetry breakdown. The superfield that we need is a boundary chiral superfield living on 
the unwarped brane, whose superpotential is $W_1(\Phi)$. To be specific, one can think of this field as of a Polonyi superfield on the Palnck brane. 
We assume  that 
$\langle \Phi \rangle^2 / M^{2}_P, \;\langle C \rangle^2 / M^{2}_P  \ll 1$. 
The low energy goldstino is given by $\eta = \frac{G_i}{\sqrt{K_{i \bar{i}}}}
\chi^i= d_i \chi^i$, where $\chi^i$ denotes a chiral fermion from the $i-th$ chiral 
supermultiplet. One finds that $|d_S|\approx |d_\Phi|
\approx 1$ and $|d_T| \approx d_{S,\;\Phi} a(\pi \rho) \; \ll \; 1$. 
This means that goldstino is the fifty--fifty mixture of dilatino and 
the Polonyi fermion. The profound contribution of the Polonyi sector 
to the supersymmetry breakdown is the price one has to pay for the vanishing 
of the tree--level cosmological constant.     
The supersymmetry breaking masses for non-Polonyi and non-moduli scalars,
$m_{i \bar{j}}^2 = g_{i \bar{j}} \mg^2 - R_{i \bar{j} k \bar{l}} G^k G\bar{l}
\mg^2$, are given by 
\beq
m_{C_2 \bar{C}_2}^2 = g_{C_2 \bar{C}_2} \mg^2 - 3 a^2 (\pi \rho) |G^\Phi|^2,
\;\;m_{C_1 \bar{C}_1}^2 = g_{C_1 \bar{C}_1} \mg^2 - 6 a^2 (\pi \rho) 
|G^\Phi|^2.
\eeq
When one goes over to the 
physical masses, then it turns out that both contributions 
to the mass squared of the scalars from the warped wall are of the order of 
$\mg^2$, whereas on the Planck wall there is a small contribution of the order 
$\mg^2 \; a^2 (\pi \rho)$ in addition to the leading term equal $\mg^2$. 
The coefficients of softly breaking supersymmetry trilinear terms 
in the scalar potential 
are of the order of unity times $\mg$ when expressed in terms of the 
canonically 
normalized scalar fields, and gaugino masses are $\approx m_{3/2}$ and universal,
due to $F^S \; \gg \; F^T$. 

Having the stage set by the detailed discussion of the RS-type warping, 
we are ready to suggest a generalization of our approach to models with  more generic 
warp factors. 
The first observation is that the vacuum shape of the warp factor  
results from the  form of the brane tensions (which are in general 
functions of brane scalars) and brane potentials 
for the bulk fields. Let us assume, that as a result of 
solving the 5d Einstein equations with the Ansatz 
\beq \label{eq:1}
ds^2 = a^2 (k R_0 |y|) ds^{2}_4 + R^{2}_0 (d x^5)^2
\eeq
we obtain a specific function $a(k R_0 |y|)$, corresponding to the BPS 
vacuum of the 5d supergravity, thus preserving four unbroken supercharges. 
As in the RS case, the 
constant $R_0$ becomes promoted to a four-dimensional field $R_0 (x)$, the 
radion, which controls the physical size of the extra dimension; $k$ is a 
dimension one constant characterizing the brane source terms. 
To obtain the effective theory, one needs to substitue the solution for 
the warp factor and bulk scalars into the action, and integrate over $x^5$.  
The terms relevant from the point of view of the radion Lagrangian 
are the Einstein--Hilbert term 
and the kinetic terms of bulk scalars. 
 
Upon reduction of 5d Einstein term one obtains some derivative terms for 
the radion, which are  part of its
kinetic Lagrangian, and the four-dimensional Einstein--Hilbert term with the 
radion-dependent coefficient
\beq
\frac{1}{2} M^{2}_P f(R_0 (x)) R^{(4)},
\eeq
where $f=2 \beta \int_0^{k \pi R_0 \rho} dy a^2 (y)$ with $\beta=\frac{M^3}{k M^{2}_P}$. 
The deduced K\"ahler potential for the radion turns out to be 
\beq K= - 3 \log f(T+\bar{T}), \label{kt} \eeq
where $f(T+\bar{T})$ is obtainded by substituting $2 k \pi \rho R_0 = T + \bar{T}$ everywhere in the function $f$. In this, lowest order, approximmation to 
the complete theory the whole information in the bulk physics is encoded 
in $f$ through the form of the warp factor.  
Already here, a comment is in order about the way the second even bulk 
scalar $V_0$, which is the real part of the 4d dilaton $S$, enters the 
K\"ahler potential. Firstly, the consequence of the fact, that it is borne 
as a member of the universal hypermultiplet, the S's K\"ahler potential 
must contain a piece $- \log (S + \bar{S})$. 
Secondly, in general the solution for the warp factor contains not only
$R_0$, but also $V_0$. An example is the warp factor in the 5d Horava--Witten 
model: $a(x^5)= (1+\alpha \sqrt{2} \frac{R_0}{V_0} (|x^5| - 
\pi \rho /2))^{1/6}$. This will necessarily lead to a mixing between $S$ and $T$ in the complete K\"ahler function. We cannot control this mixing without committing ourselves to a specific model. 

According to the above prescription the K\"ahler potential for the 5d Horava--Witten model 
is of the form 
\beq
K=-3 \log \left ( ( S + \bar{S} + T + \bar{T})^{4/3} 
- ( S + \bar{S} - T - \bar{T} )^{4/3} ) \right ),
\eeq 
and the effective superpotential has the form 
$W= 2 \sqrt{2} W_1 (1-T/S)^{1/2} - 2 \sqrt{2}  (1+ T/S)^{1/2} W_2$ (this is written with warp factor normalized to unity in the middlepoint between branes).
The well known form of these functions is recovered by replacing $T \rightarrow \alpha T$, expanding in $ \alpha$ and retaining lowest order terms.  
One can find by inspection that such a model leads to the run--away behaviour for $T$. However, a typical situation for an arbitrary form of a warp factor 
is that there exists a minimum along the direction of $T$. 

The interesting point is the presence of both $S$ and $T$ in the 
Lagrangian. 

The above reasoning illustrates, from the four-dimensional point of view, 
the important role of the warp factor: in the generic case its presence 
introduces the effective superpotential for the radion and 
leads to the stabilization of the interbrane distance {\em after} the 
breaking of supersymmetry (i.e. after switching--on $<W_1>$ and $<W_2>$). 
This is a consistent realization  of the Goldberger--Wise stabilization 
mechanism, which in addition relates stabilization to the breakdown of the 
low--energy supersymmetry. 
\vskip0.5cm
%
%
To summarize, we have identified a broad class of four-dimensional 
supergravities which are low-energy effective theories for warped 
locally supersymmetric models in five dimensions. The nontrivial warp factor 
becomes encoded in the form of the K\"ahler potential and enters the four dimensional superpotential. This in a generic case leads to stabilization of 
the radion modulus. To stabilize 
remaining moduli fields, and to make the cosmological constant vanish,
one needs to complicate and fine-tune the model, but - as shown above on 
the example of the Randall-Sundrum-class model - it is possible to achieve these goals. Further observation is that often the vacuum configuration 
of the warp factor depends not only on the radion, but also on 
the dilaton (and other moduli). 
As a result, the effective warped superpotential depends on both $S$ and $T$ moduli. 
On the other hand, the mere dependence of the superpotential on $S$ and $T$ does not guarantee stabilization - the run-away behaviour is still an option. 
We have given in some detail the analysis of the supersymmetry breaking 
in models with exponential warp factor. It is interesting to note, that 
supersymmetry breakdown is controlled there by the $F$-term of the dilaton
(and that of a Polonyi field), while the $F$-term of the $T$ modulus 
is practically zero at the vacuum.  


\section*{Acknowledgments}
This talk is based on results obtained in collaboration with 
Adam Falkowski and Stefan Pokorski.\\

We acknowledge the support by the RTN program HPRN-CT-2000-00152
and by the Polish Committee for Scientific Research grant
5~P03B~119 20 (2001-2002).

\end{document}